\documentclass[preprint,12pt]{elsarticle}

\usepackage{here}

\journal{Physica A}

\begin{document}

\begin{frontmatter}



\title{A two-layer team-assembly model\\ for invention networks}


\author[1]{Hiroyasu Inoue}
\address[1]{Osaka Sangyo University, Japan}

\begin{abstract}
Companies are exposed to rigid competition,
so they seek how best to improve the capabilities of their innovations.
One strategy is to collaborate with other companies
in order to speed up their own innovations.
Such inter-company collaborations are 
conducted by inventors belonging to the companies.
At the same time, the inventors also seem to be affected by past collaborations between companies.
Therefore, interdependency of two networks, namely inventor and company networks, exists.

This paper discusses a model that replicates two-layer networks
extracted from patent data of Japan and the United States
in terms of degree distributions.
The model replicates two-layer networks with the interdependency.
Moreover it is the only model that uses local information,
while other models have to use overall information, which is unrealistic.
In addition, the proposed model replicates empirical data
better than other models.


\end{abstract}


\end{frontmatter}



\section{Introduction}

Companies increasingly need to maximize the capacity of innovations
because of growing competition \cite{Geroski03,Czarnitzki04},
and they consider the core of that capacity to be knowledge
\cite{Grant96,Hall01,McEvily02}.
One strategy to acquire knowledge
is to collaborate with other companies
because collaborations enable companies 
to capitalize on external knowledge \cite{Chesbrough03,Laursen06}
and speed up innovations \cite{Grant96}.
Commensurate with this,
companies now place more importance on collaborations \cite{Hagedoorn03},
and the number of co-patents between companies is increasing \cite{Hicks00}.
However, companies cannot unlimitedly
acquire knowledge from other companies
since they have different corporate cultures
and unique tacit knowledge \cite{Nonaka94}.
Therefore, 
companies must have a strategy to carefully choose their collaborators.

Much research has been done on
generative models of collaboration networks
in order to understand
collaboration dynamics
\cite{Newman012,Barabasi02,Newman04,Ramasco04}.
When inventors or authors of papers collaborate,
it has been found
that 
they can create higher quality work
than those authored by solo authors \cite{Wuchty07}.
Also, inter-organizational work has more impact than
intra-organizational work \cite{Jones08}.
These previous studies focused on one-layer networks.

The collaborations in companies are 
conducted by individuals (i.e., developers or researchers) belonging to the companies.
At the same time, the inventors seem to be affected by past collaborations between companies.
Therefore, interdependency of two networks would be exist,
i.e., individual and company networks.

Here, the author proposes a model of two-layer networks,
where upper networks are expressed
by aggregations of nodes and links belonging to lower networks.
This model replicates networks
extracted from patent data of Japan and the United States
in terms of degree distributions.
Although a lot of previous studies
\cite{Jin01,Motter03,Boguna04,Gronlund04,Kimura04,Goldstein05,Li05,Chessa13,Morescalchi13}
have investigated two-layered networks,
the proposed model can replicate the observed data
better in degree distributions than those previous models.
Also, the model only uses local information,
while other models have to use overall information,
which is an unrealistic set-up in complex networks.

This paper is organized as follows.
The next section presents the data used in this study.
In section \ref{cha:model},
the model is proposed, and how it can replicate the observed networks is verified.
Finally, a summary is provided.

\section{Data}

Patents are useful for understanding what innovations occur over time
\cite{Griliches98}.
Using a massive data set enables
us to understand the tendency of innovations.
Patent data from Japan (JP) and the United States (US) are used as data sets \cite{Hall01,Tamada02} in this paper.

The identifications (IDs) of rights holders and inventors are necessary
to conduct this study.
Assigning IDs to the rights holders is easy
because their names and addresses give us sufficient information.
Companies are extracted from the rights holders based on their names.
The corporate statuses in the rights holders'
names provide the information in the JP data set.
The US data set contains information that has already been added.
In contrast, inventors are identified by name, address, and company.
The details of the process are explained in \ref{cha:invid}.

Another process is conducted
to connect each inventor with a company.
An inventor is connected to a company,
(1) if an inventor can be found on a patent applied for by a certain company,
or (2) if an inventor can be found in patents jointly applied for
by companies and there is only one common company in the companies.
Since most Japanese inventors' addresses contain names of companies,
that information is also used.

It has been more common for
teams of inventors to apply for patents,
and such patents statistically have better impact (more citations) than those authored by solo inventors \cite{Wuchty07}.
However, it is less common for
more than one company to jointly apply for patents.
This is because no company can
sell or license a patent jointly applied for
without the consent of the others.
Also, joint applications are more costly than
other solutions such as solo applications with contracts for sharing benefits,
and consequently,
joint applications between companies
are considered to be ``second best'' \cite{Hagedoorn03}.
The number of patents jointly applied for by multiple companies
account for 1.5\% of all patents in the United States and 7.8\% in Japan.

Table \ref{tbl:overview} summarizes the
fundamental data from the two data sets
with the number of patents, inventors, and companies.
Figure \ref{fig:schema} shows how
two-layer networks are created from the data sets.
The left of Figure \ref{fig:schema} shows
an example for three patents, five inventors, and three companies.
One or more inventors apply for a patent,
and each inventor works for a company.
On the basis of the tripartite network on the left,
we can create two different projected networks
for the inventors and the companies.
The inventor network is a network
where every combination of inventors has a link if they have
at least one patent in common.
The company network is defined in the same way.
If inventors who apply for a patent belong to different companies,
the companies have a link.

Figure \ref{fig:degree} plots the cumulative probability distributions of degree.
{\it Original} consists of the plots for the observed data.
The other plots are the results of models explained in a later section.
The figure plots the inventor and company networks for Japan and the U.S.
A degree is a measure to count the number of links a certain node has.
The figure indicates that
the distributions for inventors decay faster than a power law.
Previous studies found collaboration networks
have the same distributions as those in this paper \cite{Newman012,Guimera05}
or power laws \cite{Barabasi022}.
The plots of the company networks
seem to be fitted by lines, i.e., power-law distributions.
A previous study already found that the collaboration networks
of organizations in Japanese patents have power-law distributions \cite{Inoue07}.

\section{Model}

\label{cha:model}

On the basis of the observations thus far,
the author proposes a model that replicates observed networks
from the perspective of degree distributions.


This paper focuses on two-layer networks that involve
inventors and companies.
To date,
numerous generative models for networks have been developed \cite{Albert02}.
To replicate the networks in this paper,
a generative model has to
(1) explicitly assign
a group (company) to each node of a replicated network, and
(2) replicate not only a node (inventor) network but also a group network.

Gr\"onlund et al.
proposed a modified seceder model to illustrate real social networks
\cite{Gronlund04}.
Jin et al.'s model was based on the dynamics
that people actually meet \cite{Jin01}.
Bogu$\tilde{n}$\'a introduced 
the concept of social distance and found models
that could reproduce real social networks \cite{Boguna04}.
These models treat the formation of groups in observed networks
and seem similar to the model that will be proposed.
However, they create networks of individuals and
detect groups of individuals after creating individual networks
\cite{Girvan02,Newman0402,Radicchi04}.
This means groups are not explicitly given.
As previously mentioned,
the proposed model
has to explicitly provide a group to each node (item (1)).
Therefore, these studies are different from this study.

There are some models that
provide groups to nodes beforehand
when they produce networks.
Motter et al.
considered the correlation of friendships,
the positions in groups, and the correlation of positions in groups
\cite{Motter03}.
Kimura et al. demonstrated that their model
improved the prediction of real networks
by incorporating directional attachments and community structures
\cite{Kimura04}.
These models seem similar to the model that will be proposed,
but their organizational structures are given and do not grow
(item (2)).

Li and Chen also
analyzed their theoretical model that satisfies both items (1) and (2).
They showed that the degree distribution of
the model was a power law in both nodes and groups \cite{Li05}.
As explained in the previous section,
the degree distributions of the inventors are not a power law.
Therefore, their model cannot be applied either.

This section was a survey of relevant but inapplicable studies.
The following section, on the other hand,
presents two relevant models that have already been proposed and
are important as a comparison.

\subsection{Goldstein et al.'s model}

Goldstein et al. proposed a model
to replicate paper-author networks with groups of authors \cite{Goldstein05}.
Their model satisfies both items (1) and (2) in the previous section.
It is important to point out that
they did not investigate structures of networks created between groups.
Goldstein et al.'s model is comparable to the model proposed in this paper.

Figure \ref{fig:goldstein} is a diagram that the author drew to describe the model.
When a paper is created,
there is probability $\alpha$
that a new author group will be created with $N_g$ new members,
where $N_g$ is a constant.
The number of authors in the paper, $N(\lambda)$, is the first author
plus a Poisson-distributed number of additional authors.
This one-shifted Poisson distribution has parameter $\lambda$.
The probability of the one-shifted Poisson distribution, $p_{\mbox{sp}}(k)$,
is given by
\begin{equation}p_{\mbox{sp}}(k)=\frac{\lambda^{(k-1)}e^{-\lambda}}{(k-1)!},\ k=\{1,2,...\}\label{eqn:psk},
\end{equation}
where $k$ is the number of authors and $p_{\mbox{sp}}(k)$ is the probability of a paper having $k$ authors.


If no new group is created,
an existing author group is chosen using the following probability distribution:
\begin{equation}
p_{g}(q)=\frac{q}{N_{p}},\label{eqn:pg}
\end{equation}
where $q$ is the number of papers
that this group has published, $N_{p}$ is the total number of papers
in the network,
and $p_{g}(q)$ is the probability of an existing group having authored a paper.

When adding each author,
there is probability $\beta$
to choose an author from another group.
After choosing a group,
a selection of the author is done by using another preferential process.
The probability of selecting author $i$ in the group is
\begin{equation}
  p_{\mbox{a}}(i)=\frac{k_i+1}{\sum{k_j+N_g}},\label{eqn:pa}
\end{equation}
where $k_i$ is the number of papers written by author $i$,
$\sum{k_j}$ is the sum of the number of authorships
of authors in the group, and $N_g$ is the number of
authors in the group.

Goldstein et al.'s model is simple and comparable to
the model the author will propose.
However, it is to be noted that
Goldstein et al.'s model requires overall information
to calculate Eq. (\ref{eqn:pg}) and (\ref{eqn:pa}).
The availability of overall information is
normally unrealistic because collaboration networks are vast and complex.

\subsection{Guimera et al.'s model}

\label{cha:guimera}

The model proposed later is based on Guimera et al.'s model \cite{Guimera05},
which aims to replicate the self-assembly of creative teams
and has two parameters, which are of the fraction of newcomers
in new productions ($p$) and the tendency of incumbents
to repeat previous collaborations ($q$).

Figure \ref{fig:guimeraProcess} outlines
the process of how the model progresses.
The model has an endless pool of newcomers.
Newcomers become incumbents after being selected.
The model adds members to a team according to $m$\footnote{
There are various ways of creating the sequence for $m$:
e.g., keep $m$ constant,
or draw $m$ from the observed distribution.
The latter is used in this paper.
}.
Probability $p$ indicates a member drawn from the pool of incumbents.
If a member has already been chosen from the pool of incumbents
and there is already another incumbent that is already connected but has not been chosen,
a new member is chosen with probability $q$ from the incumbents.
Otherwise, a member is chosen from all the incumbents.
The process is repeated $m$ times for each team.

\subsection{The proposed model}

\label{cha:prmodel}

The purpose of this paper is to propose a new model
based on Guimera et al.'s model
that can replicate the two-layered networks obtained from the empirical data
better than the models studied previously.
Figure \ref{fig:lambdaProcess} outlines the proposed model.
Guimera et al.'s model remains at the top left.
The model contains a new process for 
choosing companies (X) and 
creating companies (Y).
There is a branch when an inventor is a newcomer.
If the inventor is the first member of a team, Y is executed.
If it is not, X is executed.
X has a parameter, $r$.
Here, $r^k$, where $k$ is the number of
companies already included in the patent,
is the probability of choosing a company
from the pool of all existing companies.
Then, the newcomer or the incumbent is assigned to the chosen company.
If $r^k$ is not true,
the same company that one of the members already belongs to
is chosen for the newcomer or the incumbent.
Y has a parameter, $s$, which
is the probability of creating a new company.
Then, the newcomer is assigned to the chosen company.
If $s$ is not true, a company is randomly chosen
from the pool of all existing companies,
and then the newcomer is assigned to it.

\subsection{Simulation results and discussion}

The author applied
Goldstein et al.'s, Guimera et al.'s, and the proposed models
to replicate two-layer networks observed in the empirical data
in order to see how the proposed model improves on the replication compared to previous models.
The comparison was conducted after tuning parameters of each model.
The tuned parameters were
$\alpha$, $\beta$, $\lambda$, and $N_g$ for Goldstein et al.'s model,
$p$ and $q$ for Guimera et al.'s model,
and $p$, $q$, $r$, and $s$ for the proposed model.

The tuning was conducted through {\it simulated annealings} \cite{Karkpatrick83}.
Parameters are initially set according to values
that seem to be the closest values that can be obtained from the observed data.

The initial values of the parameters in Goldstein et al.'s model were
$\alpha=0.02$, $\beta=0.17$, $\gamma=1.53$, and $N_g=26$ for Japanese data
and 
$\alpha=0.05$, $\beta=0.28$, $\gamma=0.6$, and $N_g=9$ for the U.S. data.
The sizes of steps to search neighborhoods in simulated annealings were
0.01 for $\alpha$ and $\beta$, 0.1 for $\gamma$, and 1 for $N_g$.
The following observed values were used as initial values.
The probability where new companies are found in patents was used as $\alpha$.
The average fraction of another company's inventor was used as $\beta$.
Since $\gamma$ is a parameter of the one-shifted Poisson distribution Eq. (\ref{eqn:psk}),
$\gamma$ can be obtained through the least squares method to fit the one-shifted Poisson distribution to the distribution of the number of authors.
The average of the number of authors in patents was used as $N_g$.

Two parameters, $p$ and $q$, which are necessary to run Guimera et al.'s model, were initially set as
$p=0.73$ and $q=0.69$ for Japanese data
and
$p=0.78$ and $q=0.66$ for the U.S. data.
The sizes of steps were
0.01 for $p$ and $q$.
The author calculated the fraction of newcomers to members in every patent
and the average of the fractions was used as $p$.
The author also calculated the fractions of repeated collaborations to all collaborations in every patent,
and then the average of the fractions was used as $q$.
Each inventor was randomly assigned to a company
from a pool in the simulation.
The pool had the same number of companies as the observed data.

New parameters in the proposed model, namely, $r$ and $s$,
were initially set as $r=0.06$ and $s=0.09$ for Japanese data
and
as $r=0.05$ and $s=0.07$ for the U.S. data.
The sizes of steps were
0.01 for $r$ and $s$.
The author obtained the probability distribution
that a newcomer belongs to a different company other than $k$ companies already included in the patent.
Then, $r$ was set through the least squares method for the distribution.
$s$ was set to the probability that a new inventor belongs to a new company.
The other initial settings were the same as
those in the simulations of Guimera et al.'s model.

The simulated annealing created and evaluated the networks 1,000 times.
A probability to adopt worse parameters than the current parameters exponentially decayed as the repetitions progress.
Every repetition of the simulated annealing
created 1,696,635 patents to replicate Japanese networks
and 722,350 patents to replicate the U.S. networks.

To evaluate the replicated networks,
Kolmogrov-Smirnov (KS) statistic, which indicates distances of two cumulative probability distributions was used.
Here, two cumulative probability distributions are drawn from the obtained and replicated networks.
Since there are two different networks to evaluate, i.e., inventor and company networks,
the sum of KS statistics of the two networks was used as an evaluation value.

Table \ref{tbl:KSSA} lists the results of the parameter fitting.
The Total column indicates the sums of KS statistics of inventor and company networks,
which were used as evaluation values in the simulated annealings.
The KS statistics are given for the inventor's and company's cumulative degree distributions for Goldstein et al.'s,
Guimera et al.'s, and the proposed models.
The smaller a KS statistic is, the closer two distributions drawn from a replicated and the observed networks are.

The total values of Guimera et al.'s model in Table \ref{tbl:KSSA}
show that the model is not comparable to the others.
The fitted parameters are 
$p=0.92$ and $q=0.93$ for Japanese data
and
$p=0.77$ and $q=0.68$ for the U.S. data.
Although Guimera et al.'s model was able to replicate the inventor's network well,
it has large KS statistics in company networks.
Figure \ref{fig:degree} also shows large deviations from the original data in company networks.
Since all inventors are randomly assigned to companies in the simulations,
similar numbers of inventors are assigned to all companies.
Therefore, the company distribution does not match the observed one.

On the other hand,
Goldstein et al.'s model
seems to be comparable to the proposed model.
The fitted parameters are 
$\alpha=0.18$, $\beta=0.07$, $\lambda=1.8$, and $N_g=20$ for 
Japanese data
and
$\alpha=0,06$, $\beta=0.18$, $\lambda=2.0$, and $N_g=2$ for 
the U.S. data.
The evaluation value in the JP data is superior to the one in the proposed model
(Table \ref{tbl:KSSA}).
However,
we can see large deviations from the observed data
in the inventors' network in Figure \ref{fig:degree}.
This is because
KS statistics are absolute values, and they evaluate the tails of distributions less.
This large deviation in the tail shows that Goldstein et al.'s model does not seem to be a better model than the proposed model.

The proposed model appears to be able to replicate the observed networks better,
although the proposed model does not always show better performance than Goldstein et al.'s model in the KS statistics.
It is clear from Figure \ref{fig:degree} that
the proposed model does not have as large a deviation as the other models.
The fitted parameters are 
$p=0.64$, $q=0.65$, $r=0.10$, and $s=0.25$ for Japanese data
and
$p=0.43$, $q=0.60$, $r=0.03$, and $s=0.28$ for the U.S. data.
Note that the proposed model does not require overall information, which Goldstein et al.'s model requires.
Since overall information is unrealistic,
this is one point of improvement of the proposed model.

Since the model replicates the degree distributions of inventors and companies
better than the other models,
the characteristics of the proposed model
may help to understand the mechanism for choosing the partners as inventors and companies.
The following three characteristics can be deduced from the proposed model.
(1) Inventors with many connections to other inventors
have greater possibilities of obtaining other connections in the future;
since the results of fitting showed $p$ and $q$ have large values,
the path of $p$ and $q$ (Figure \ref{fig:lambdaProcess}) often happens.
Therefore, an inventor with many links is likely to be involved in a team.
(2) Companies with many inventors can acquire inter-company connections;
a new connection between companies can mainly be obtained
from the path where $p$ is true, $q$ is false, and $r^k$ is true.
Since an incumbent is randomly chosen in the process,
a company with many incumbents is likely to be chosen.
(3) Inter-company connections grow
by attracting new connections to the existing inter-company connections;
as it has already been mentioned in (1),
the path of $p$ and $q$ often happens.
If there is an inter-company team, other inventors tend to be involved in the team.
Therefore, inter-company links are likely to increase.

It has to be admitted that
the proposed model does not truly replicate the observed data
because it cannot pass statistically strict tests, such as a KS test.
However, no other models, even for one-layer networks, seem to be able to replicate true distributions either.
Therefore, the proposed model can be considered as the first step toward a better model.

\section{Summary}

This paper attempted to clarify interdependency
between inventor and company networks
using patent data from Japan and the United States.
Also, two different networks were created from tripartite graphs of patents, companies, and inventors.

The author created a model 
to replicate two-layer networks
to understand the interdependent evolution of the networks.
The model is based on Guimera et al.'s model
and was able to replicate the observed networks better in terms of cumulative degree distributions
than other models.
A key characteristics of the proposed model is
that all processes only use local information, which is not achieved by other models.

\bibliographystyle{model1-num-names}
\bibliography{paper}

\clearpage


\appendix

\section{Results of inventor identification}
\label{cha:invid}

Assigning IDs to inventors requires an additional process for the patent data.
The original patent data did not have IDs.
A comprehensive study has been done on this identifying process \cite{Trajtenberg06},
which considers names, addresses, affiliations, co-inventors, technological classifications, citations, and different spellings of names.
However,
this detailed process mainly aims to net out the movements of inventors.
This paper does not consider the movements.
Thus, it is sufficient to identify 
inventors by their names, addresses, and affiliations.

Figure \ref{fig:invaccprob} plots
cumulative probabilities of the number of patents per inventor.
The Japanese and U.S. data have similar patterns.
After the identification,
4,649,617 names in patents were merged into 1,806,259 inventors in Japan
and 4,301,229 names merged into 1,923,241 inventors in the U.S.

\newpage

\renewcommand{\thefigure}{\arabic{figure}}
\renewcommand{\thetable}{\arabic{table}}

\begin{table*}[H]
\caption{Overview of data sets:
The range of years in which patents were applied for is labeled ``Duration.''
The table lists the numbers of patents and companies that are included in the patents.
Inventors working in companies were extracted.}
\begin{center}
{\footnotesize
\label{tbl:overview}
\begin{tabular}{l|rr|rr}
 & \multicolumn{2}{l|}{US} & \multicolumn{2}{l}{JP} \\
\hline
Duration (year) & 1963-1999 & & 1994-2008 & \\
Total number of patents & 2,923,922 & & 1,967,361 & \\
Number of companies & 33,515 & & 72,841 & \\
Number of inventors in companies & 285,418 & & 829,052 & \\
Number of patents by multiple inventors & 347,450 & & 1,043,639 & \\
Total number of patents by companies & 722,350 & & 1,696,635 & \\
Number of patents by multiple companies & 28,345 & & 132,704 & \\
\end{tabular}
}
\end{center}
\end{table*}

\begin{table*}[H]
\caption{Kolmogorov-Smirnov (KS) statistics after parameter fittings given by simulated annealings:
Each number is a KS statistic between distributions of a model and the observation.
A small KS statistic means a distribution is close to the observed distribution.\label{tbl:KSSA}}
\begin{center}
{\footnotesize
\begin{tabular}{l||r|r|r}
JP & \multicolumn{1}{l}{Inventor\hspace*{2ex}} & \multicolumn{1}{l}{Company} & \multicolumn{1}{l}{Total\hspace*{5ex}} \\
\hline
\hline
Goldstein et al.'s model & 0.05 & 0.12 & 0.17 \\
\hline
Guimera et al.'s model & 0.04 & 0.93 & 0.97 \\
\hline
Proposed model & 0.18 & 0.03 & 0.21 \\
\hline
\end{tabular}

\begin{tabular}{l}
\\
\end{tabular}

\begin{tabular}{l||r|r|r}
US & \multicolumn{1}{l}{Inventor\hspace*{2ex}} & \multicolumn{1}{l}{Company} & \multicolumn{1}{l}{Total\hspace*{5ex}} \\
\hline
\hline
Goldstein et al.'s model & 0.05 & 0.65 & 0.70 \\
\hline
Guimera et al.'s model & 0.03 & 0.98 & 1.01 \\
\hline
Proposed model & 0.23 & 0.13 & 0.46 \\
\hline
\end{tabular}
}
\end{center}
\end{table*}

\clearpage

\begin{figure*}[H]
\begin{center}
\includegraphics[scale=0.5]{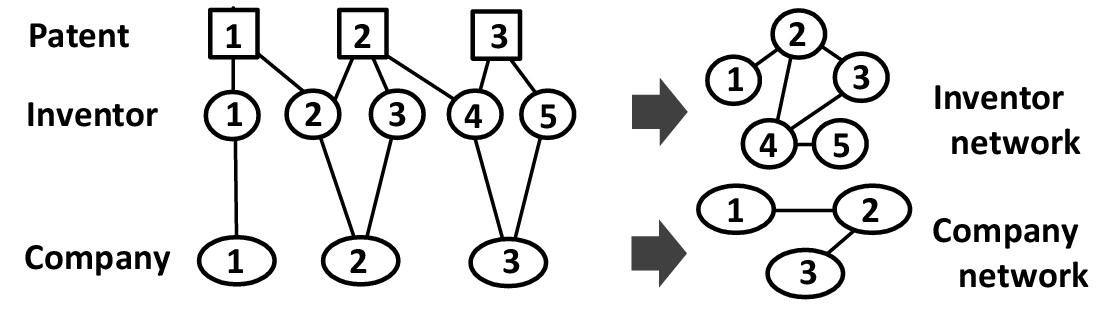}
\caption{Process to create two-layer networks from observed data: The figure shows how two different networks were created from patent data. A patent can be applied for by two or more inventors. Also, a patent can be applied for by two or more companies. For example, patent 1 has two inventors, 1 and 2. These inventors are connected in the inventor network. The inventor respectively belong to companies 1 and 2. Therefore, companies 1 and 2 are connected.\label{fig:schema}}
\end{center}
\end{figure*}

\clearpage

\begin{figure*}[b]
   \begin{minipage}{0.4\hsize}
     \begin{center}
       \includegraphics[scale=0.4]{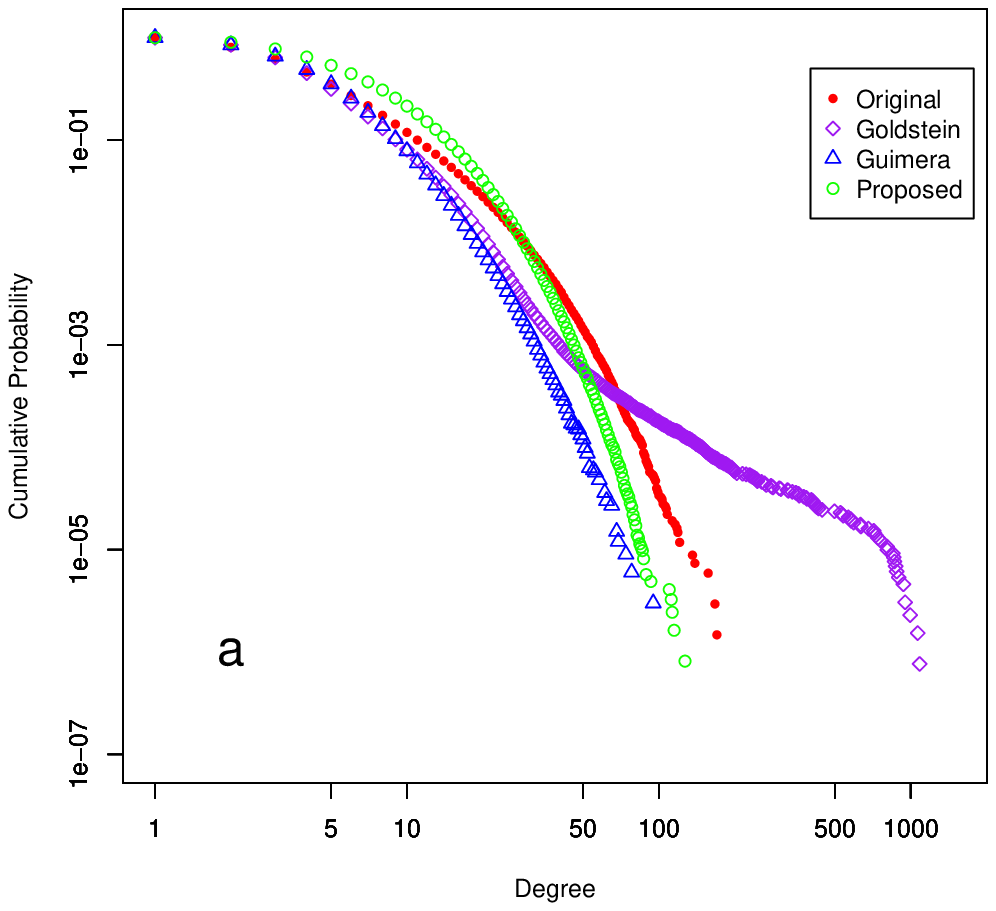}
     \end{center}
   \end{minipage}
   \hspace{12ex}
   \begin{minipage}{0.4\hsize}
     \begin{center}
       \includegraphics[scale=0.4]{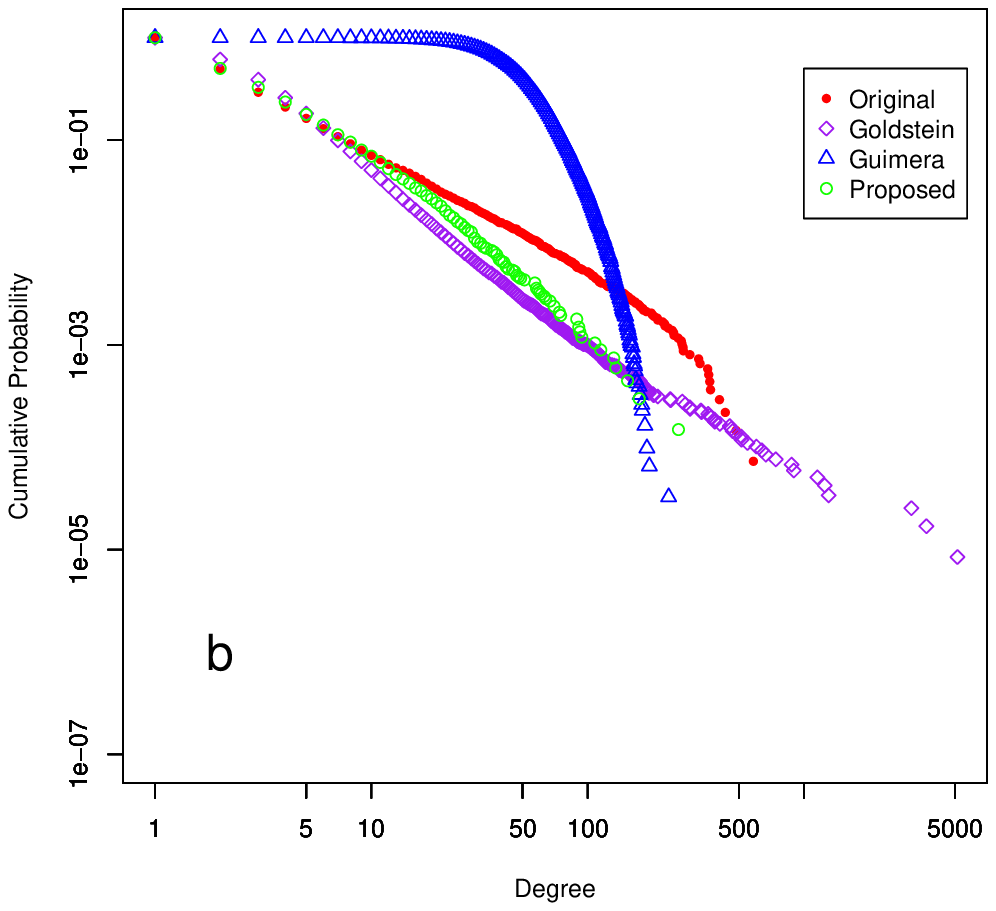}
     \end{center}
   \end{minipage}
\\
   \begin{minipage}{0.4\hsize}
     \begin{center}
       \includegraphics[scale=0.4]{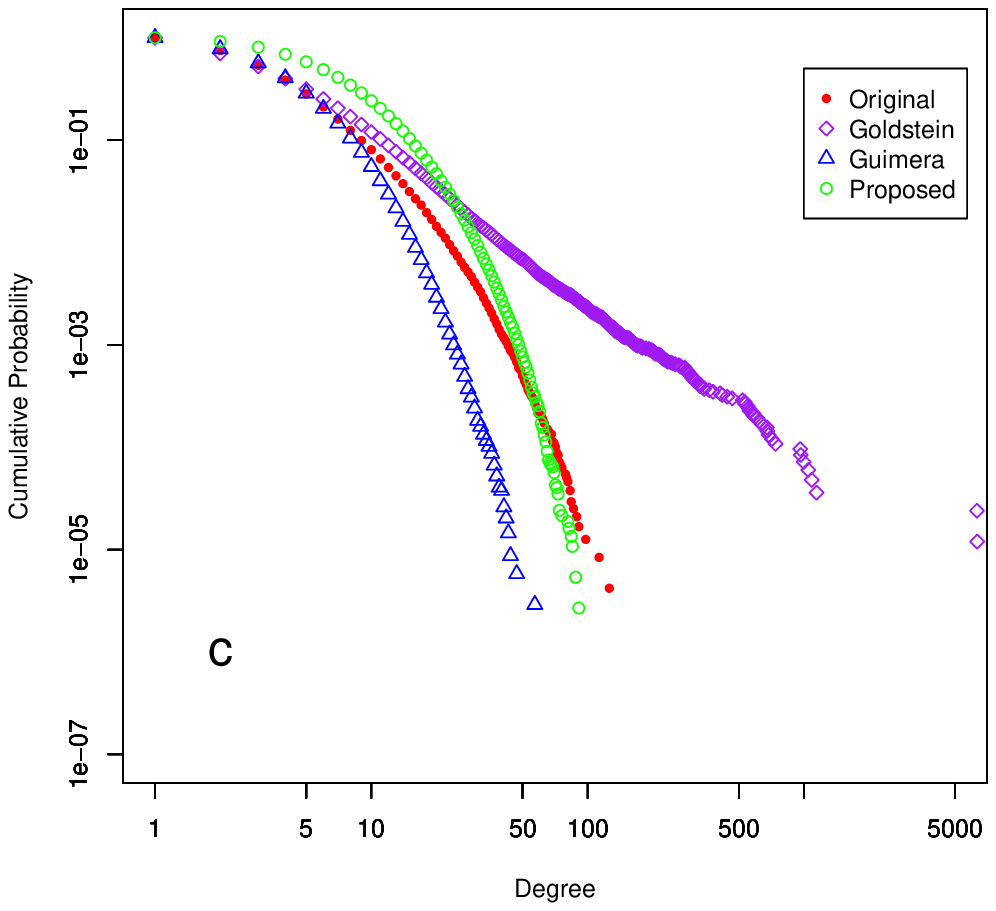}
     \end{center}
   \end{minipage}
   \hspace{12ex}
   \begin{minipage}{0.4\hsize}
     \begin{center}
       \includegraphics[scale=0.4]{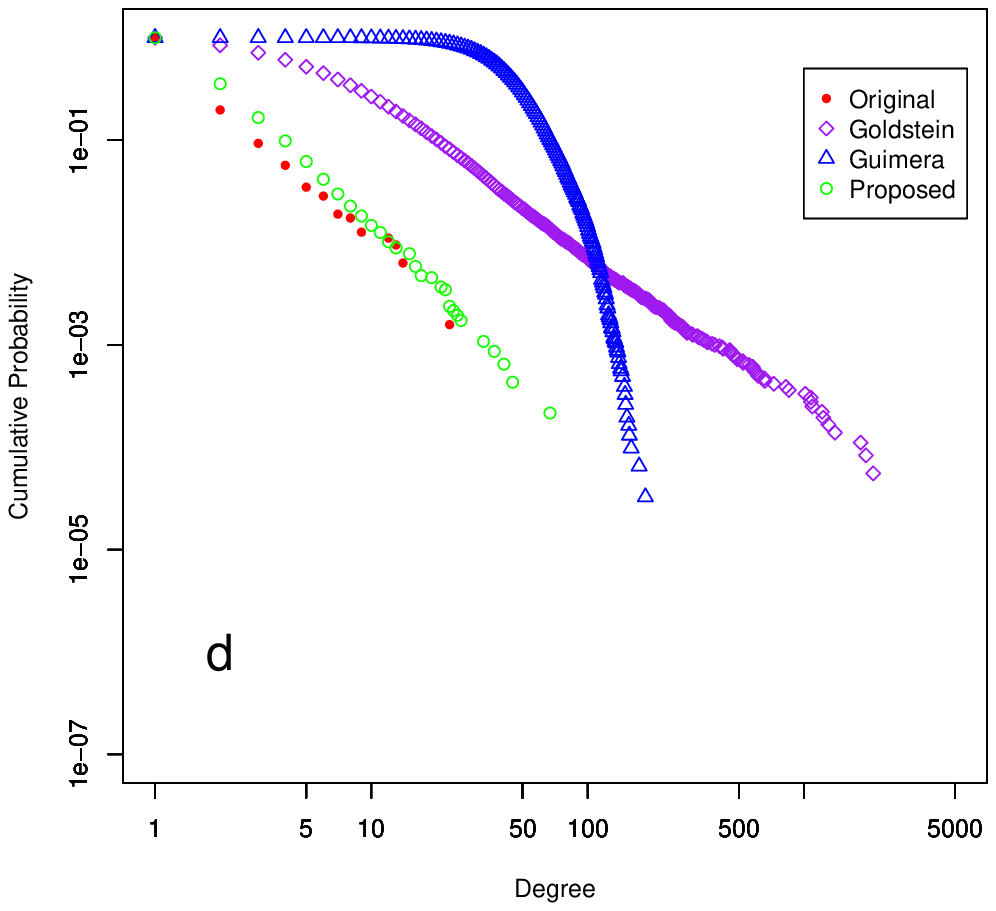}
     \end{center}
   \end{minipage}
   \caption{Cumulative probability distributions for degree of inventors' and companies' networks. {\it Original} is the observed data. {\it Goldstein}, {\it Guimera}, and {\it Proposed} are the data derived from each model with parameters obtained from the simulated annealing. {\bf a}: Japanese inventors {\bf b}: Japanese companies {\bf c}: U.S. inventors {\bf d}: U.S. companies.
\label{fig:degree}}
\end{figure*}

\clearpage

\begin{figure*}[H]
\begin{center}
\includegraphics[scale=0.7]{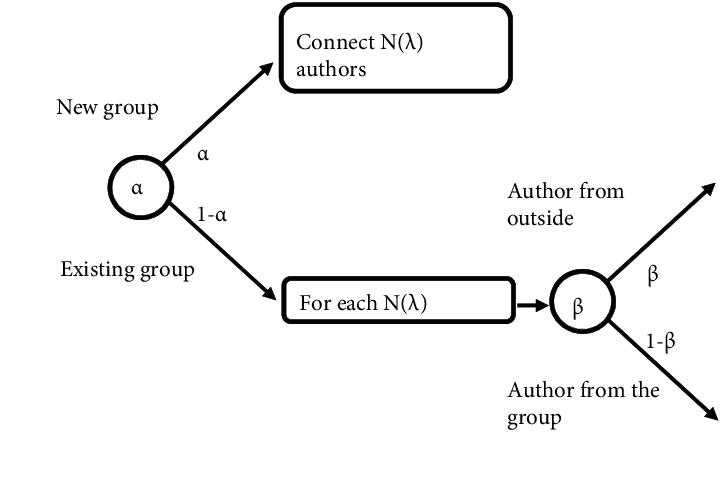}
\caption{Goldstein et al.'s model \protect \cite{Goldstein05}: The process is repeated for every paper (i.e., patent in this paper). With probability $\alpha$, a new group is created. The group has $N(\lambda)$ members. In other cases, an existing group is chosen. With probability $\beta$, an author is chosen from other groups. In other cases, an author is chosen from the group already chosen. In the choice, the author will be chosen in proportion to the number of times the author has been chosen. \label{fig:goldstein}}
\end{center}
\end{figure*}

\begin{figure*}[H]
\begin{center}
\includegraphics[scale=0.5]{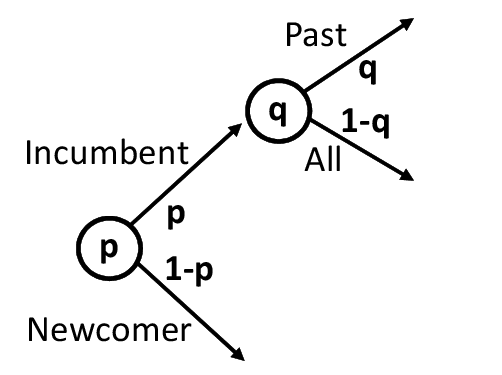}
\caption{Process in Guimera et al.'s model: The process is repeated by the number of members in a team. An incumbent is randomly chosen with probability $p$. If $p$ is not true, a newcomer is created. After $p$ is true, $q$ is tested. With probability $q$, an incumbent is a past collaborator of team members. However, if $q$ is not true, an incumbent is randomly chosen from all incumbents.\label{fig:guimeraProcess} \protect \cite{Guimera05}}

\end{center}
\end{figure*}

\begin{figure*}[H]
\begin{center}
\includegraphics[scale=0.5]{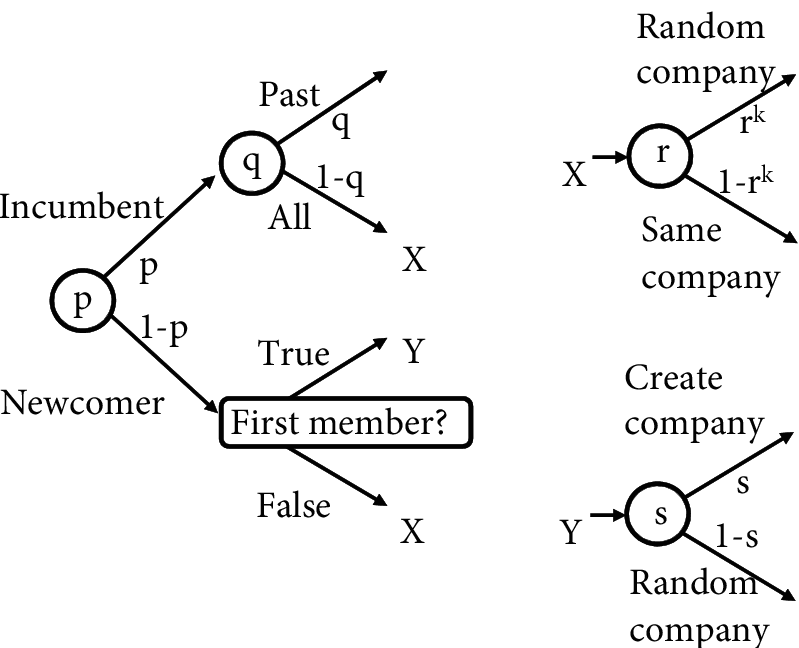}
\caption{Process for proposed model: It has Guimera et al.'s model at the top left. X and Y on the left indicate jumps to X and Y on the right.\label{fig:lambdaProcess}}
\end{center}
\end{figure*}

\clearpage

\setcounter{figure}{0}
\setcounter{table}{0}

\renewcommand{\thefigure}{A.\arabic{figure}}
\renewcommand{\thetable}{A.\arabic{table}}

\begin{figure*}[H]
\begin{center}
\includegraphics[scale=0.7]{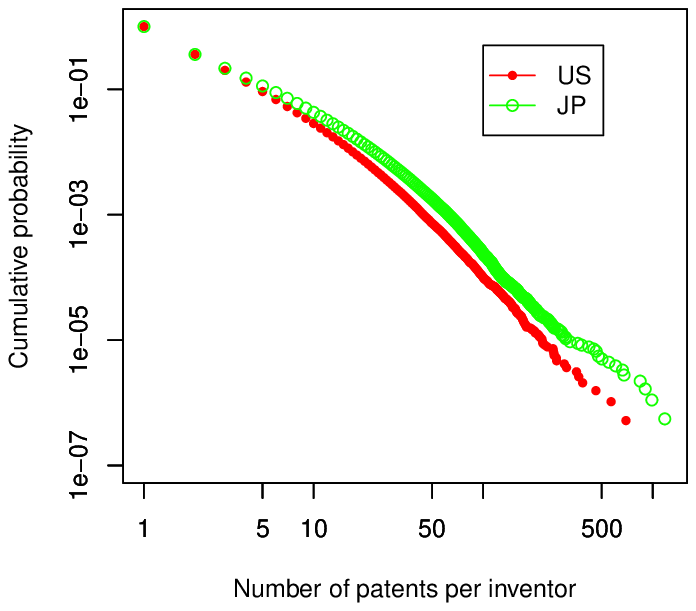}
\caption{Cumulative probability of number of patents per inventor: The horizontal axis plots the number of patents per inventor and the vertical axis plots the cumulative probability. Japan and the U.S. have similar patterns.\label{fig:invaccprob}}
\end{center}
\end{figure*}





\end{document}